\documentclass[journal]{IEEEtran}

\usepackage{multirow}
\usepackage[table,xcdraw]{xcolor}
\usepackage{graphicx}
\usepackage{amsmath}
\usepackage{amsfonts}
\usepackage{algorithm}
\usepackage{amsthm,amssymb}
\usepackage{mathrsfs}
\usepackage{algorithmic}
\usepackage{makecell}
\usepackage{easyReview}
\usepackage{booktabs}
\setcellgapes{3pt}
\usepackage{verbatim}
\setcellgapes{3pt}

\usepackage[justification=centering]{caption}
\usepackage{color}
\usepackage[subfigure]{tocloft}
\usepackage{subfigure}
\usepackage{tabularx}
\usepackage{amsmath, bm}
\usepackage{hyperref}
\usepackage{subfigure}
\usepackage{siunitx}
\hypersetup{hidelinks,
	colorlinks=true,
	allcolors=blue,
	pdfstartview=Fit,
	breaklinks=true}

\usepackage{cite}

\ifCLASSINFOpdf
\else
\fi

\begin{document}
	\title{FAS-LLM: Large Language Model–Based Channel Prediction for OTFS-Enabled Satellite–FAS Links}
	
\author{Halvin~Yang,~\IEEEmembership{Member,~IEEE, }%
		Sangarapillai~Lambotharan,~\IEEEmembership{Senior Member,~IEEE, }%
		Mahsa~Derakhshani,~\IEEEmembership{Senior Member,~IEEE}%
		 \thanks{Halvin Yang and Sangarapillai Lambotharan are with the Institute for Digital Technologies, Loughborough University London, Loughborough University, London, UK (email: h.yang6@lboro.ac.uk, s.lambotharan@lboro.ac.uk)}
		\thanks{Mahsa Derakhshani is with the Wolfson School of Mechanical Electrical and Manufacturing Engineering
			at Loughborough University, Loughborough, U.K. (e-mail: M.Derakhshani@lboro.ac.uk).} 
}

\markboth{Submitted for Review}%
{Shell \MakeLowercase{\textit{et al.}}: Bare Demo of IEEEtran.cls for IEEE Journals}

\maketitle 

\begin{abstract}
This paper proposes FAS-LLM, a novel large language model (LLM)–based architecture for predicting future channel states in Orthogonal Time Frequency Space (OTFS)-enabled satellite downlinks equipped with fluid antenna systems (FAS). The proposed method introduces a two-stage channel compression strategy combining reference-port selection and separable principal component analysis (PCA) to extract compact, delay–Doppler–aware representations from high-dimensional OTFS channels. These representations are then embedded into a LoRA-adapted LLM, enabling efficient time-series forecasting of channel coefficients. Performance evaluations demonstrate that FAS-LLM outperforms classical baselines including GRU, LSTM, and Transformer models, achieving up to 10\,dB normalized mean squared error (NMSE) improvement and threefold root mean squared error (RMSE) reduction across prediction horizons. Furthermore, the predicted channels preserve key physical-layer characteristics, enabling near-optimal performance in ergodic capacity, spectral efficiency, and outage probability across a wide range of signal-to-noise ratios (SNRs). These results highlight the potential of LLM-based forecasting for delay-sensitive and energy-efficient link adaptation in future satellite IoT networks.
\end{abstract}


%
\IEEEpeerreviewmaketitle
\section{Introduction}\label{sec:intro}

\subsection{Background}
The coming decade will witness the deployment of tens of billions of Internet-of-Things devices (IoTDs)—sensor tags, asset trackers, precision-agriculture nodes, and wearables—often located in oceans, deserts, and polar routes where terrestrial backhaul is unavailable. Recent GSMA reports indicate that more than 60\% of the planet’s surface remains outside reliable cellular coverage, and in those regions fewer than 20\% of installed IoTDs achieve continuous connectivity~\cite{GSMA_IoT_Maps}. Low-Earth-orbit (LEO) satellite constellations are emerging as the most scalable non-terrestrial network (NTN) solution \cite{lin2021leo6g}: unlike high-altitude platforms (HAPs) or long-endurance unmanned aerial vehicles (UAVs), LEO satellites provide permanent global coverage without persistent energy or air-traffic costs~\cite{alfattani2023haps}.

Due to the high orbital velocities of LEO satellites, Doppler shifts of several kHz (S-band) to tens of kHz (Ka-band) are induced, far exceeding those observed in terrestrial macrocells~\cite{shi2022leo}. Such large Doppler spreads result in rapidly time-varying channels that severely degrade the performance of conventional modulation schemes like OFDM, which rely on quasi-static channel assumptions over each symbol duration. In particular, the loss of orthogonality between subcarriers introduces inter-carrier interference (ICI), where energy from one subcarrier leaks into adjacent subcarriers \cite{jeon1999equalization}. This leads to corrupted symbol detection, reduced spectral efficiency, and increased receiver complexity due to the need for more sophisticated equalization. Together with the increased channel estimation burden, these effects cause significant performance degradation when standard OFDM receivers are applied in high-Doppler scenarios.

To overcome these challenges, Orthogonal Time–Frequency Space (OTFS) modulation was developed. OTFS maps symbols to a two-dimensional delay–Doppler (DD) grid, allowing each data symbol to experience the full diversity of the channel rather than a single fading instance~\cite{hadani2017otfs,raviteja2019embedded}. This transforms the doubly-selective channel into a nearly time-invariant representation across an OTFS frame, significantly improving robustness to Doppler effects. Channel estimation then reduces to recovering a small set of dominant DD taps, rather than tracking thousands of rapidly time-varying subcarrier gains.

On the device side, massive IoT applications impose stringent constraints on size, power consumption, and hardware complexity. To address these requirements, the fluid antenna system (FAS) has emerged as a promising architecture. An FAS employs a single RF chain connected to a set of ports distributed along a linear or surface-based structure, such as a microfluidic cavity or conductive rail, where only one port is active at a time~\cite{wong2021fluid}. The system dynamically selects the optimal port based on the instantaneous channel conditions, thereby achieving spatial diversity without the need for multiple RF chains.

However, since the ports are typically spaced by fractions of a wavelength to maintain a compact device footprint, their channel responses tend to be highly correlated. This presents a significant challenge for port selection, as the gain difference between candidate ports can be minimal and highly sensitive to small channel variations \cite{wong2020limits}. Furthermore, switching between ports incurs non-negligible delay and control overhead \cite{tong2025hardware}, meaning the optimal port index must be known in advance to maintain continuous data reception. As a result, accurate channel prediction is critical—not only to support proactive link adaptation at the satellite transmitter but also to enable timely and reliable port reconfiguration at the IoTD receiver.

\subsection{Related Work}
Orthogonal Time--Frequency Space (OTFS) modulation has demonstrated strong potential across diverse use cases, including multiuser communications~\cite{chen2025otfs} and large-scale MIMO systems~\cite{mehrotra2024sparse}. Its adaptability is further evidenced by successful integrations with cutting-edge technologies such as reconfigurable intelligent surfaces (RIS)~\cite{li2022hybridRIS}, non-orthogonal multiple access (NOMA)~\cite{ding2019otfsnoma}, and chirp-based waveforms for joint communication and sensing~\cite{zegrar2024otfschirp}. These studies reinforce OTFS's suitability as a foundational waveform for advanced wireless systems.

However, this versatility comes at the cost of significant computational and energy burdens, particularly for low-power IoT devices (IoTDs)~\cite{xiao2022otfs}. The implementation of key OTFS operations---most notably, the two-dimensional inverse symplectic finite Fourier transform (ISFFT)~\cite{Mohammed2021} and associated channel estimation and equalization procedures~\cite{lakew2023intelligent}---demands considerable processing resources. These constraints have driven recent research efforts toward complexity reduction and improved energy efficiency. Notable examples include low-complexity equalizer designs~\cite{surabhi2020lowcomplexity} and transceiver architectures optimized for energy-limited IoT scenarios. Yet, scalability remains an ongoing concern, with lightweight implementations still required to accommodate dense and heterogeneous IoT deployments~\cite{sui2023low}.

Fluid antenna systems (FAS) have recently emerged as a promising technology due to their ability to reconfigure spatial positions dynamically, enabling enhanced signal reception and link robustness without relying on conventional channel estimation or digital beamforming~\cite{wong2021fluid,new2024fluid}. This makes them inherently well suited to environments with limited computational resources or stringent energy budgets.

Analytical models of FAS have been further refined to account for practical factors such as receiver noise~\cite{yang2023performance}, allowing more accurate evaluation of their performance under realistic conditions. Beyond basic connectivity, FAS has been investigated for simultaneous data and energy transfer in integrated systems, where its switching behavior can be leveraged to optimize trade-offs between spectral and energy efficiency~\cite{lin2024performance,zhang2024joint}.

The applicability of FAS extends beyond traditional communication use cases. For instance, recent research has integrated FAS with index modulation techniques to increase spatial encoding capacity~\cite{yang2024position}, and with integrated sensing and communication (ISAC) architectures to support dual-purpose signal processing~\cite{wang2024fluid}. This combination of configurability, energy efficiency, and computational simplicity positions FAS as a compelling solution for future wireless networks, particularly in scenarios involving device mobility or constrained IoT hardware.

In recent years, deep learning has become a powerful tool in wireless communications, particularly for problems where analytical models struggle to capture the underlying complexity~\cite{maatouk2024llmtelecom}. While its early applications have focused on areas such as beam selection and resource optimization~\cite{zhang2024beamforming}, the concept of data-driven port selection in fluid antenna (FA) systems remains relatively underexplored. Port switching in FAs, especially in mobile scenarios, is inherently a temporally dynamic task, requiring not just spatial awareness but also predictive insight into how channel conditions evolve over time~\cite{li2024fluid}.

Traditional sequence models such as LSTM and GRU~\cite{greff2017lstm,chung2014gru} have been applied to similar time-dependent tasks, yet their limited depth and representation capacity often hinder performance in high-dimensional or rapidly changing environments. These models can struggle to extrapolate under non-stationary conditions, which are common in mobile channels with frequent fading and port correlation effects.

More recently, the wireless research community has begun to examine the potential of transformer-based architectures, including large language models (LLMs), for physical-layer tasks. Though initially developed for natural language processing, their ability to capture long-term dependencies and process structured sequences at scale makes them compelling candidates for modeling wireless channel dynamics \cite{zhou2024llmsurvey}. Preliminary work has shown promising results in using LLMs for tasks like energy efficiency \cite{liu2025llmris} and task offloading \cite{zhou2025generative}.

Building upon the exploration of transformer-based architectures, recent studies have begun to harness the potential of LLMs for channel prediction. Liu \textit{et al.} \cite{Liu2024LLM4CP} introduced an LLM-driven framework for channel state information (CSI) prediction, leveraging the model's inherent sequence learning capabilities to improve multi-step forecasting accuracy. Similarly, \cite{Fan2025CsiLLM} proposed the CSI-LLM model, which extends LLM architectures for downlink channel prediction, effectively aligning the design with sequential modelling tasks typical of NLP. Complementing these efforts, Jin \textit{et al.} developed LinFormer \cite{Jin2025LinFormer}, a lightweight transformer variant optimized for time-aware MIMO channel prediction, achieving superior performance with reduced complexity. Collectively, these works illustrate the growing impact of LLM-based methods in advancing wireless channel modelling, particularly in dynamic and high-mobility scenarios.

To date, the only known study to explore fluid antenna systems in satellite communication contexts is our prior work, which analyzed the performance limits of OTFS-based satellite links with fluid antennas. However, that work  did not address the challenges of real-time port selection or predictive operation. Moreover, while recent research has explored deep learning and LLMs for high-level wireless tasks such as beam prediction and CSI modeling, no existing work has attempted to apply large language models to physical-layer channel prediction in the delay--Doppler--port space of OTFS-based FAS satellite links. This paper addresses that gap by proposing a novel LLM-powered predictor designed to capture the evolving delay–Doppler characteristics and port-dependent variations of satellite-to-ground channels with fluid antenna receivers.

\subsection{Motivation and Contributions}
Existing channel prediction methods fall short in addressing the compound challenges posed by LEO satellite mobility, OTFS modulation complexity, and the switching constraints of fluid antenna systems, particularly when deployed on energy-constrained IoT terminals. To fill this gap, we introduce FAS-LLM, the first large language model tailored for delay--Doppler channel prediction in OTFS-enabled satellite links with fluid antennas. The key novelties and contributions of our work are summarised as follows:

\begin{itemize}
	\item \textbf{Novel application of LLMs for delay--Doppler--port channel prediction}: While prior work has applied deep learning to CSI forecasting, our work is among the first to explore the use of large language models for predicting physical-layer channels in the delay--Doppler domain. Crucially, we extend this to include port-wise spatial variation in fluid antenna systems—an integration not previously addressed in the literature.

	\item \textbf{Joint delay--Doppler and port selection under realistic switching constraints}: We uniquely model and predict both the delay--Doppler response and the optimal FAS port one frame in advance, explicitly accounting for practical switching delays. This enables real-time implementation of FAS in realistic NTN settings where instantaneous switching is infeasible.
	
	\item \textbf{Novel two-stage CSI compression pipeline}: We propose a novel reference-port filtering stage followed by separable PCA, which reduces the input size by over 99\% while preserving the channel’s dominant energy structure. This novel compression strategy allows high-dimensional CSI to be effectively processed by the LLM without exceeding token limits.
	
	\item \textbf{Comprehensive benchmarking and superior performance}: We conduct extensive comparisons against existing prediction schemes, including LSTM, GRU, and Kalman-RNN models. Our FAS-LLM achieves significantly lower error and higher communication performance across capacity and outage metrics, validating its practical superiority.
	
	\item \textbf{Scalability to massive IoT in non-terrestrial networks}: By enabling proactive and low-complexity port switching on a single-RF-chain receiver, our approach addresses the pressing need for scalable, energy-efficient connectivity in global NTN-based IoT deployments. This unlocks new avenues for massive device access in regions beyond terrestrial coverage.
\end{itemize}

The remainder of the paper presents the system model in Section II, followed by a detailed description of the FAS‑LLM architecture and its training methodology in Section III. Section IV provides extensive simulation-based performance evaluations to demonstrate the effectiveness of the proposed approach. Finally, Section V concludes the paper with a summary of key insights and future research directions.

\section{System Model} 

\begin{figure}[t]  
	\centering
	\includegraphics[width=0.85\linewidth]{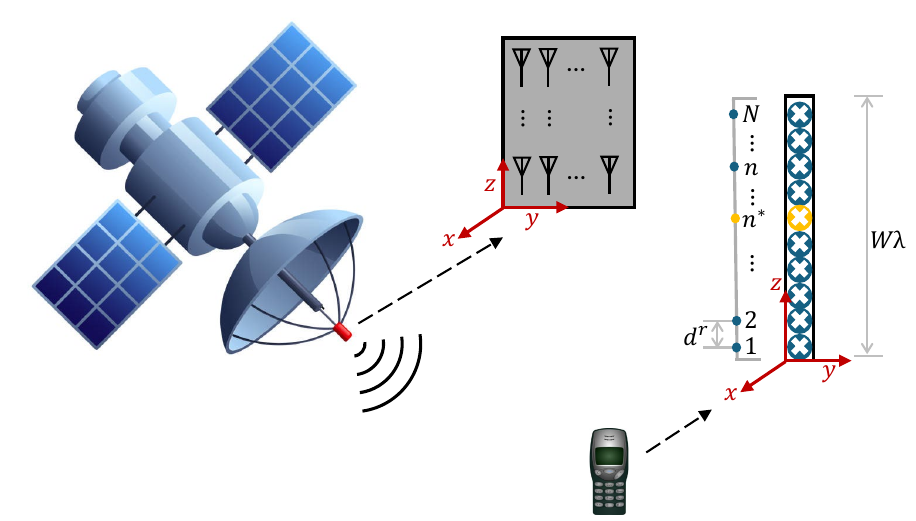}
	\caption{System model of a downlink Satellite-FAS channel.}
	\label{fig:SystemModel}
	\vspace{-10pt}
\end{figure}

Figure \ref{fig:SystemModel} illustrates the LEO-satellite downlink considered in this paper.  
The signal departs from a uniform planar array (UPA) on the satellite, propagates through a Ricean delay–Doppler
channel and is captured by a single fluid antenna located at the receiver.  
We now develop the mathematical model for each stage in this downlink system.

\subsection{Satellite Transmit Array and OTFS Waveform}
\label{sec:TxArray}

The satellite carries a UPA comprising
$N_t = N_x N_y$ elements arranged on an $N_x \times N_y$ rectangular grid
with half-wavelength spacing $d_t = \lambda/2$. 
Let $x_{t_x,t_y}(t)$ denote the continuous-time baseband waveform radiated
from transmit element $(t_x,t_y)$,  where $t_x \in {1,...,N_x}$ and $t_y \in {1,...,N_y}$. Under OTFS modulation, this waveform is
\begin{align}
	x_{t_x,t_y}(t)
	&= \sum_{k=0}^{M-1} \sum_{\ell=0}^{N_\nu-1}
	X_{t_x,t_y}[k,\ell]\;
	g_{\mathrm{tx}}\!\bigl(t-kT\bigr)\,
	e^{\,j2\pi \ell \Delta f (t-kT)},
\end{align}
where $\Delta f$ is the subcarrier spacing, $T = 1/\Delta f$ is the symbol
duration, $g_{\mathrm{tx}}(\cdot)$ is the transmit pulse, and
$X_{t_x,t_y}[k,\ell]$ is the TF-domain data symbol assigned to that element.

\subsection{Fluid-Antenna Receiver and Spatial Correlation}
\label{sec:FAScorr}

At the receiver, we consider a fluid-antenna system (FAS) composed of $N_p$ ports arranged along a linear rail. The ports are uniformly spaced by $d_r \ll \lambda$ along the $z$-axis, resulting in strong spatial correlation due to the sub-wavelength aperture. This linear arrangement enables fine-grained sampling of the spatial domain with minimal hardware complexity.

To model spatial correlation across the FAS, we assume planar wavefronts impinging at an elevation angle $\theta$. Under this assumption, the spatial correlation matrix between ports $n_p$ and $n_p'$ is given by 
\begin{align}
	\bigl[R_{\mathrm{FAS}}\bigr]_{n_p,n_p'}
	&= J_0\!\left(2\pi\,\frac{d_r}{\lambda}\,|n_p-n_p'| \sin\theta\right),
\end{align}
where $J_0(\cdot)$ is the zeroth-order Bessel function of the first kind. This closed-form expression reflects the underlying structure of a 1D uniform linear array (ULA) receiving plane waves in free space.

To ensure numerical stability during Cholesky factorisation, we compute the square root of the correlation matrix as $R_{\mathrm{FAS}}^{1/2} = \operatorname{chol}(R_{\mathrm{FAS}} + \epsilon I_{N_p})$, where $\epsilon \ll 1$ is a small diagonal loading constant, $I_{N_p}$ is the identity matrix of size $N_p$ and $\operatorname{chol}$ is the Cholesky decomposition.

\subsection{Delay–Doppler Channel Model}
\label{sec:DDchannel}

Each transmit element illuminates the FAS receiver through a narrowband Ricean channel comprising a deterministic line-of-sight (LoS) path and $P$ scattered multipath components. The time-varying impulse response is discretised into delay and Doppler bins. Sampling the channel at delays $m_\tau \in \{0,\dots,M_\tau{-}1\}$ and Doppler shifts $n_d \in \{0,\dots,N_\nu{-}1\}$ yields the discrete delay–Doppler coefficient $H_{n,t_x}[n_d,m_\tau](q)$, where $q \in \mathbb{Z}_{\ge 0}$ denotes the OTFS frame index.

The full channel is modelled as: %
\begin{align}
	H_{n_p,t_x}[n_d,m_\tau](q)
	= \sqrt{\frac{\kappa}{\kappa + 1}}\;
	b_n\,a_t\,
	e^{\,j 2\pi \nu_{\mathrm{LoS}} q T_{\mathrm{fr}}}\,
	\delta_{n_d,\nu_{\mathrm{LoS}}}\,
	\delta_{m_\tau,0}
	\nonumber\\
	\quad
	+ \sqrt{\frac{1}{\kappa + 1}}
	\sum_{p=0}^{P-1}
	\sqrt{\rho_p}\;
	e^{\,j 2\pi \nu_p q T_{\mathrm{fr}}}\,
	\bigl[R_{\mathrm{FAS}}^{1/2} \mathbf{z}_p\bigr]_n\,
	g_{p,t_x}\,
	\delta_{n_d,\nu_p}\,
	\delta_{m_\tau,\tau_p}.
	\label{eq:FullPathModel}
\end{align}

Here, $\kappa$ is the Rice factor controlling the relative strength of the LoS path. We assume isotropic transmit elements, so $a_t = 1$ for all $t$. The receive-side spatial signature of the LoS path is captured by $b_n = [R_{\mathrm{FAS}}^{1/2}]_{n,1}$, the $n$-th entry of the dominant eigenvector of the correlation matrix. For each scattered path $p$, the tuple $(\tau_p, \nu_p, \rho_p)$ specifies the delay, Doppler shift, and average power, respectively, typically drawn from a 3GPP delay–Doppler tap profile.


The receive and transmit spatial responses of each scatterer are modelled as independent complex Gaussian vectors: $\mathbf{z}_p \sim \mathcal{CN}(\mathbf{0}, \tfrac{1}{2} I_{N_p})$ for the FAS ports, and $g_{p,t} \sim \mathcal{CN}(0, \tfrac{1}{2})$ per UPA element. The OTFS frame duration $T_{\mathrm{fr}}$ determines the time increment between successive snapshots. 

In~\eqref{eq:FullPathModel}, the LoS term is pre‑beamformed at the satellite, resulting in uniform transmit gains $a_t = 1$. The spatial response at the FAS is expressed by $b_n = [R^{1/2}_{\text{FAS}}]_{n,1}$, where the Bessel‑law correlation in $R_{\text{FAS}}$ captures the elevation angle of arrival (AoA) $\theta$. For each scattered NLoS path, the tuple $(\tau_p, \nu_p, \rho_p)$ determines its delay–Doppler bin, while the vector $R^{1/2}_{\text{FAS}} \mathbf{z}_p$ imposes the AoA‑dependent spatial progression across ports. Transmit‑side steering toward the angle of departure (AoD) is represented by the independent coefficients $g_{p,t}$. Thus, AoA effects are embedded in the receive correlation matrix, while AoD manifests in the element‑specific gains $g_{p,t}$. Beamforming renders the LoS path rank‑one, whereas NLoS paths maintain full transmit diversity via $g_{p,t}$.

This model captures key aspects of the physical propagation. Each scatterer contributes to a single DD bin $(\tau_p, \nu_p)$, and its influence rotates frame-by-frame via a Doppler-induced phase factor $e^{j2\pi \nu_p q T_{\mathrm{fr}}}$. The FAS correlation matrix determines how power is spread across receive ports via the term $R_{\mathrm{FAS}}^{1/2} \mathbf{z}_p$.

Finally, we assemble the full delay–Doppler channel tensor by stacking coefficients over all ports $n_p = 1,\dots,N_p$, transmit elements $t_x = 1,\dots,N_t$, Doppler bins $n_d$, and delay bins $m_\tau$: %
\begin{align}
	H(q)
	&= \Bigl\{ H_{n_p,t_x}[n_d,m_\tau](q) \Bigr\}
	\;\in\; \mathbb{C}^{N_p \times N_t \times N_\nu \times M_\tau}.
\end{align}
This tensor forms the OTFS D-D Channel block and serves as the input to the predictive compression architecture described in the next section.

\subsection{Two-Stage Channel Compression}

To efficiently feed channel matrices into a transformer-based LLM, it is essential to reduce both their dimensionality and redundancy. The full reference-port matrix $H_{\rm ref}(q) \in \mathbb{C}^{N_t \times (M_\tau N_\nu)}$ contains $N_t M_\tau N_\nu$ complex coefficients, which far exceed the context window size of most LLM architectures. Note that choosing a single reference port eliminates the dependency of the LLM input on the number of ports $N_p$, since the sub-wavelength spacing causes their channels to differ only by a deterministic phase shift, allowing the full tensor to be regenerated analytically after prediction without increasing compression complexity. Moreover, a significant fraction of these coefficients represent redundant or low-energy components. To address this, we propose a two-stage compression strategy: (i) \emph{reference-port selection} to eliminate spatial redundancy across fluid antenna ports, followed by (ii) \emph{separable principal component analysis (PCA)} to exploit low-rank structure in both spatial and delay–Doppler domains.

\subsubsection*{Reference-Port Selection}

In fluid antenna systems (FAS), the receive ports are densely arranged along a sub-wavelength rail. As a result, their individual channel responses differ only by deterministic phase shifts, while the underlying multipath structure remains invariant. Specifically, each port observes identical delay–Doppler taps, modulated by port-dependent phase ramps. This observation allows one to extract the full channel matrix from a single port and reconstruct the others analytically (more about reconstruction in subsection \ref{subsection:reconstruct}).

The reference port is selected as the one exhibiting the highest average received power across a training set of $N_{\text{train}}$ frames: 
\begin{align}
	n_p^\star 
	&= \arg\max_{n_p=1,\dots,N_p}
	\frac{1}{N_{\mathrm{train}}}
	\sum_{q=1}^{N_{\mathrm{train}}}
	\bigl\lVert H(q)[n_p,:,:,:]\bigr\rVert_{F}^{2}.
\end{align}
In practice, the correlation among ports is nearly uniform when $d_r \ll \lambda$, allowing us to fix $n_p^\star = 1$ by convention. The extracted matrix for frame $q$ is then given by %
\begin{align}
	H_{\mathrm{ref}}(q)
	&= H(q)\bigl[n_p = n_p^\star\bigr]
	\;\in\;\mathbb C^{N_t\times(M_\tau N_\nu)}.
\end{align}

\subsubsection*{Separable PCA-Based Compression}\label{section:PCA}

Even after selecting a single reference port, $H_{\mathrm{ref}}(q)$ remains too large for direct LLM input. To further reduce dimensionality, we perform \emph{separable PCA} to independently compress the spatial and delay–Doppler dimensions. The basis matrices are derived empirically from the training set by computing the sample covariance matrices: %
\begin{align}
	R_s
	&= \frac{1}{N_{\mathrm{train}}}
	\sum_{q=1}^{N_{\mathrm{train}}}
	H_{\mathrm{ref}}(q)\,H_{\mathrm{ref}}(q)^{H}
	\;\in\;\mathbb C^{N_t\times N_t},\\
	R_d
	&= \frac{1}{N_{\mathrm{train}}}
	\sum_{q=1}^{N_{\mathrm{train}}}
	H_{\mathrm{ref}}(q)^{H}\,H_{\mathrm{ref}}(q)
	\;\in\;\mathbb C^{(M_\tau N_\nu)\times(M_\tau N_\nu)}.
\end{align} 
Eigendecomposing these yields %
\begin{align}
	R_{s} &= U_{s}\,\Lambda_{s}\,U_{s}^{H}, 
	&\quad
	A_{s} &= \bigl[u_{s,1}\;\cdots\;u_{s,r_{s}}\bigr]
	\;\in\;\mathbb C^{N_{t}\times r_{s}},\\
	R_{d} &= U_{d}\,\Lambda_{d}\,U_{d}^{H}, 
	&\quad
	A_{d} &= \bigl[u_{d,1}\;\cdots\;u_{d,r_{d}}\bigr]
	\;\in\;\mathbb C^{(M_\tau N_\nu)\times r_{d}}.
\end{align} 
where $A_s$ and $A_d$ consist of the leading eigenvectors that capture the majority of energy in each dimension.

\subsubsection*{Rank Selection Criteria}

The ranks $r_s$ and $r_d$ denote the number of retained principal components in the spatial and delay–Doppler domains, respectively. They are selected to capture a desired fraction (typically 90–95\%) of total signal energy: %
\begin{itemize}
	\item \textbf{Spatial rank $r_s$:} %
	\begin{align}
		\frac{\sum_{i=1}^{r_s} \lambda_{s,i}}{\sum_{i=1}^{N_t} \lambda_{s,i}} \;\ge\; 0.90.
	\end{align}
	Empirically, for $N_t \approx 8$ under moderate scattering, $r_s$ lies in the range of 3 to 5.
	
	\item \textbf{Delay–Doppler rank $r_d$:} %
	\begin{align}
		\frac{\sum_{i=1}^{r_d} \lambda_{d,i}}{\sum_{i=1}^{M_\tau N_\nu} \lambda_{d,i}} \;\ge\; 0.90.
	\end{align}
	Although $M_\tau N_\nu$ may be on the order of 1024, it is often observed that 16–32 components suffice to preserve over 90\% of the total energy.
\end{itemize}

\subsubsection*{Low-Dimensional Projection and Vectorisation}

Given $A_s$ and $A_d$, each reference-port matrix $H_{\mathrm{ref}}(q)$ is projected into a compressed form: %
\begin{align}
	C(q)
	&= A_{s}^{H}\,H_{\mathrm{ref}}(q)\,A_{d}
	\;\in\;\mathbb C^{r_{s}\times r_{d}}.
\end{align}
The resulting matrix $C(q)$ contains $r_s r_d$ principal-component coefficients. These represent the dominant spatial and delay–Doppler modes of the channel. For typical configurations, $r_s r_d$ lies between 50 and 100—orders of magnitude smaller than $N_t M_\tau N_\nu \sim 8000$.

We then vectorise $C(q)$ as: %
\begin{align}
	\mathbf{c}(q) &= \mathrm{vec}\bigl(C(q)\bigr)
	\;\in\;\mathbb{C}^{r_s r_d},
\end{align}
and feed the resulting vector—after optional delta encoding and quantisation—into the LLM, accompanied by appropriate frame and port meta-tokens. This drastically reduces the sequence length, focusing model capacity on the most informative aspects of the channel.


In summary, \emph{reference-port selection} eliminates spatial redundancy by exploiting the deterministic phase relationships among densely packed FAS ports. \emph{Separable PCA compression} further reduces dimensionality by leveraging the low-rank structure in both spatial and delay–Doppler domains. Together, these steps constitute a memory- and compute-efficient front-end, enabling practical LLM-based channel prediction without sacrificing key statistical characteristics.

\subsection{Channel Reconstruction from the Predicted Code}
\label{subsection:reconstruct}
Given a predicted code vector from the LLM at time frame \( q+1 \), denoted \( \widehat{\mathbf{c}}(q+1) \in \mathbb{C}^{r_s r_d} \), the full delay–Doppler channel tensor is reconstructed through three deterministic steps: (i) unvectorisation of the code into a matrix form, (ii) projection into the spatial and delay–Doppler subspaces using learned PCA bases, and (iii) deterministic spatial replication across fluid-antenna ports via a phase ramp.

\subsubsection*{Step 1: Unvectorisation}

The first step reshapes the predicted vector into a low-rank coefficient matrix: %
\begin{align}
	\widehat{C}(q+1)
	&= \mathrm{unvec}_{\,r_s \times r_d}\bigl(\widehat{\mathbf{c}}(q+1)\bigr)
	\;\in\; \mathbb{C}^{r_s \times r_d}.
\end{align}
This matrix represents the principal component coefficients in the reduced spatial and delay–Doppler domains.

\subsubsection*{Step 2: Expansion in the PCA Subspaces}

The compressed matrix \( \widehat{C}(q+1) \) is then mapped back to the reference-port delay–Doppler domain using the spatial and delay–Doppler basis matrices \( A_s \in \mathbb{C}^{N_t \times r_s} \) and \( A_d \in \mathbb{C}^{M_\tau N_\nu \times r_d} \), which were obtained via separable PCA over a training set defined in Section~\ref{section:PCA}. The expansion is given by %
\begin{align}
	\widehat{H}_{\mathrm{ref}}(q+1)
	&= A_s \, \widehat{C}(q+1) \, A_d^{H}
	\;\in\; \mathbb{C}^{N_t \times (M_\tau N_\nu)}.
\end{align}
This operation is fully deterministic, does not require iterative optimization, and reconstructs the reference-port channel by projecting the compressed code back into the original signal space.

\subsubsection*{Step 3: Spatial Replication via Deterministic Phase Ramp}

To recover the full four-dimensional channel tensor across all fluid-antenna ports, we exploit the deterministic phase relationships induced by the receive geometry. The ports lie on a uniform linear array aligned along the $z$-axis (see Fig.\,\ref{fig:SystemModel}b), with inter-port spacing \( d_r \ll \lambda \). Under the standard plane-wave model, a wave arriving from elevation angle \( \theta \) induces a per-port phase shift given by %
\begin{align}
	\Phi_{n_p}
	&= \exp\left(-j\,2\pi\,(n_p-1)\,\tfrac{d_r}{\lambda} \sin\theta\right),
	\qquad n_p = 1,\dots,N_p.
\end{align}
This phase factor is constant across transmit elements, delay bins, and Doppler bins because the relative path geometry remains identical for all components.

The full reconstructed channel tensor is thus obtained by applying this phase offset element-wise across ports: %
\begin{align}
	\widehat{H}(q+1)[n_p, t_x, n_d, m_\tau]
	&= \widehat{H}_{\mathrm{ref}}(q+1)[t_x, n_d, m_\tau] \cdot \Phi_{n_p}.
\end{align}
Equivalently, the tensor may be expressed compactly as %
\begin{align}
	\widehat{H}(q+1) &= \boldsymbol{\Phi} \odot \widehat{H}_{\mathrm{ref}}(q+1), \\
	\boldsymbol{\Phi} &= 
	\bigl[\Phi_1, \dots, \Phi_{N_p}\bigr]^{T}
	\otimes \mathbf{1}_{N_t \times N_\nu \times M_\tau},
\end{align}
where $ \odot $ denotes elementwise multiplication and $ \mathbf{1}_{N_t \times N_\nu \times M_\tau} $ is a tensor of ones used to replicate the phase vector across the remaining dimensions.

Since the phase ramp is fully determined by the array geometry and arrival angle $ \theta $, the final reconstruction step incurs no additional computational complexity beyond a single complex multiplication per receive port. The entire procedure—unvectorisation, basis expansion, and deterministic spatial replication—efficiently maps the LLM’s compressed output back into the full channel tensor $ \widehat{H}(q+1) \in \mathbb{C}^{N_p \times N_t \times N_\nu \times M_\tau} $ required for subsequent processing at the receiver.

\section{Methods}
This section introduces the LLM training procedure, tokenisation details, and evaluation of the predictive compression pipeline.

\begin{figure*}[htbp]
	\centering
	\includegraphics[width=16cm]{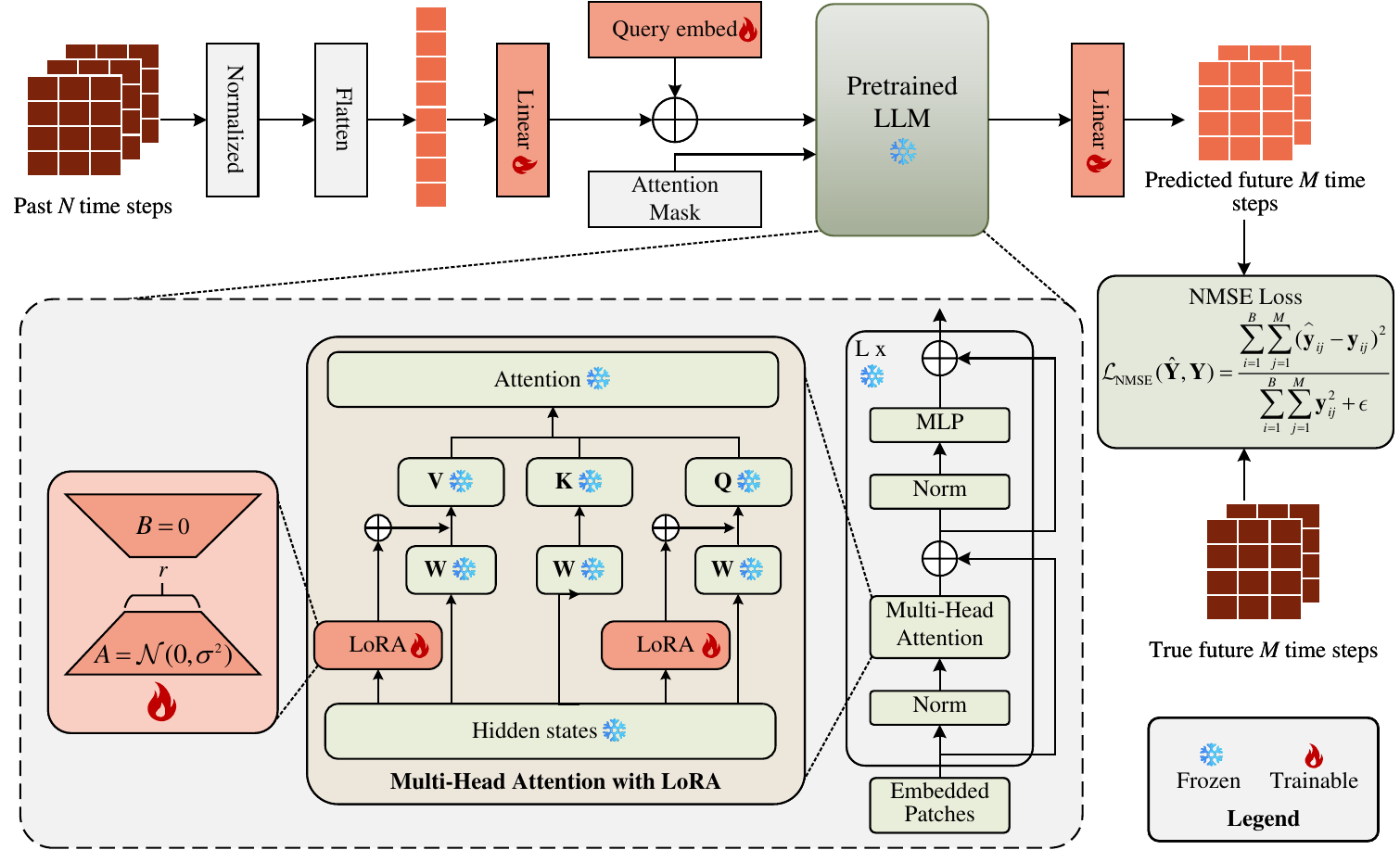}
	\caption{The network design of the proposed FAS-LLM framework.}
	\label{fig:FAS-LLM}
\end{figure*}

\subsection{Architecture of FAS-LLM} 
As shown in Fig.~\ref{fig:FAS-LLM}, we propose a time-series prediction framework built upon a pre-trained LLM, adapted using LoRA \cite{devalal2018lora} to efficiently model complex temporal dependencies in high-dimensional sequential data. The model is designed to predict future representations based on past observations. Specifically, given an input $ \mathbf c(q) = \mathbf{X} \in \mathbb{R}^{B \times N \times D} $ representing the past $ N $ time steps of a $ D $-dimensional feature array in the $B$th mini-batch - the number of independent training sequences processed simultaneously, the goal is to predict $ M $ future time steps $ \hat{\mathbf{Y}} \in \mathbb{R}^{B \times M \times D} $. The processing pipeline is as follows:

First, each input feature vector at every time step is normalized and flattened. Then, a linear projection is applied to map the input into the model's internal embedding space: 
\begin{equation}\label{eq:LLM1}
\mathbf{X}_{\text{emb}} = \mathbf{X} \mathbf{W}_{\text{in}} + \mathbf{b}_{\text{in}}, \quad \mathbf{W}_{\text{in}} \in \mathbb{R}^{D \times D}.
\end{equation} 

Simultaneously, a set of learnable query embeddings $ \mathbf{Q} \in \mathbb{R}^{1 \times N \times D} $ is initialized to predict the future representations. These queries are embedded through a linear layer and concatenated with the embedded past features:
\begin{equation}\label{eq:LLM2}
\mathbf{Z}_0 = \left[ \mathbf{X}_{\text{emb}},\ \mathbf{Q} \right] \in \mathbb{R}^{B \times (M+N) \times D}.
\end{equation} 

The backbone of the LLM is a multi-layer Transformer architecture, which excels at modelling temporal dynamics through its self-attention mechanism. Within each Transformer block, the Multi-Head Self-Attention (MHSA) operation allows every token, including future queries, to attend to all past and current tokens, facilitating both short- and long-range dependency modelling: 
\begin{equation}\label{eq:LLM3}
    \text{Attention}(\mathbf{Q}, \mathbf{K}, \mathbf{V}) = \text{softmax}\left( \frac{\mathbf{Q} \mathbf{K}^\top}{\sqrt{d_k}} \right) \mathbf{V},
\end{equation}
where $ \mathbf{Q}, \mathbf{K}, \mathbf{V} \in \mathbb{R}^{B \times (N+M) \times D_h} $ are derived from linear projections of $ \mathbf{Z}_0 $, and $ D_h $ is the per-head dimension.

After processing through the LLM, the final $ M $ embeddings corresponding to the query tokens are extracted: %
\begin{equation}\label{eq:LLM4}
    \mathbf{H}_{\text{pred}} = \mathbf{Z}_L[:, -M:, :] \in \mathbb{R}^{B \times M \times D_h}.
\end{equation}

These are then projected back to the original feature space using an output linear layer: %
\begin{equation}\label{eq:LLM5}
    \hat{\mathbf{Y}} = \mathbf{H}_{\text{pred}} \mathbf{W}_{\text{out}} + \mathbf{b}_{\text{out}}, \quad \mathbf{W}_{\text{out}} \in \mathbb{R}^{D_h \times D}.
\end{equation}

Overall, by combining the expressive power of LLMs, the temporal abstraction capability of Transformer blocks, and the lightweight efficiency of LoRA, this framework provides a scalable and accurate solution for forecasting in complex temporal modelling scenarios.
\subsection{Training of the FAS-LLM}  
To enhance parameter efficiency during fine-tuning, LoRA is applied to the attention projection matrices.  
LoRA is a technique designed to efficiently fine-tune large pretrained models while keeping most of the model's parameters frozen, and is particularly beneficial for models like Transformers, where fine-tuning can be computationally expensive due to the large number of parameters. As shown in Fig. \ref{fig:FAS-LLM}, rather than updating the full model, LoRA inserts trainable low-rank matrices into specific layers (e.g., query and value projections), modifying the weight update as \cite{devalal2018lora}: %
\begin{equation}\label{eq:LLM6}
    \Delta\mathbf{W} = \alpha \cdot \mathbf{A} \mathbf{B}, \quad \mathbf{A} \in \mathbb{R}^{D_h \times r},\ \mathbf{B} \in \mathbb{R}^{r \times D_h},
\end{equation}
where $ r \ll D_h $ and $ \alpha $ is a scaling factor. This strategy significantly reduces the number of trainable parameters while preserving expressive capacity, making it highly suitable for temporal tasks requiring continual adaptation.

The training process employs the Normalized Mean Squared Error (NMSE) loss function to evaluate the model's prediction accuracy. The NMSE loss is computed as follows: %
\begin{equation}\label{eq:LLM7}
    \mathcal{L}_\text{NMSE}(\mathbf{\hat{Y}}, \mathbf{Y}) = \frac{\sum_{i=1}^{B} \sum_{j=1}^{M} (\hat{\mathbf{y}}_{ij} - \mathbf{y}_{ij})^2}{\sum_{i=1}^{B} \sum_{j=1}^{M} \mathbf{y}_{ij}^2 + \epsilon},
\end{equation}
where $\mathbf{Y}$ represents the true $ M $ future time steps. $ \mathbf{y}_{ij} $ refers to the $ j $-th value in the $ i $-th batch of target data, which represents the actual feature value for the $ j $-th future time step in the sequence. The loss function compares the predicted values $ \hat{\mathbf{y}}_{ij} $ with the corresponding target values $ \mathbf{y}_{ij} $, and calculates the squared difference between them. 

The NMSE loss is normalized by the sum of squared target values, $ \mathbf{y}_{ij}^2 $, to ensure stability in the loss function, especially when the target values are close to zero. A small constant $ \epsilon $ is added to the denominator to prevent division by zero or numerical instability. This loss function provides a clear evaluation of the model’s performance, effectively measuring the accuracy of temporal predictions while considering the variability of target data.
Assuming that $\mathcal{D}$ denotes the dataset and $\mathbf{\theta}$, the training process is summarized \textbf{Algorithm \ref{alg:Training}}.

\begin{algorithm}
    \caption{Training Process for FAS-LLM}
    \label{alg:Training}
    \begin{algorithmic}[1]
        \REQUIRE $\mathcal{D}$.
        \ENSURE Trained model parameters $\bm{\theta}$
        \FOR{each training epoch}
        \FOR{each batch $(\mathbf{X}, \mathbf{Y})$ from $\mathcal{D}$ }
            \STATE{Obtain the internal embedding $\mathbf{X}_\text{emb}$ using Eq. (\ref{eq:LLM1}).}
            \STATE{Predict the $M$ future time steps $\mathbf{\hat{Y}}$ according to Eqs. (\ref{eq:LLM3}-\ref{eq:LLM5}).}
            \STATE{Compute NMSE loss using Eq. (\ref{eq:LLM7}).}
            \STATE{Backpropagate and update model parameters $\bm{\theta}$ with the optimizer.}
        \ENDFOR
    \ENDFOR
    \end{algorithmic}
\end{algorithm}
\section{Experimental Setup and Simulation Results}
This section details the simulation dataset, experimental configurations, and evaluation results. All simulations are performed on a high-performance computing server equipped with an Intel Xeon CPU (2.3 GHz, 256 GB RAM) and two NVIDIA RTX 4090 GPUs (each with 24 GB SGRAM), utilizing the PyTorch framework to implement the proposed models.

\subsection{Experimental Settings}

\subsubsection{Network and Training Parameters}  
We adopt the LLaMA-3-1B model \cite{grattafiori2024llama} as the pretrained large language model (LLM), which features a hidden dimension of $ D_h = 2048 $. For LoRA-based fine-tuning, the rank reduction factor is set to $ r = 8 $. The model is optimized using the AdamW algorithm. The learning rate is initialized at $ 1 \times 10^{-3} $, with a batch size of 64 and a total of 100 training epochs. The forecasting task is defined with a fixed past window length of $ N = 50 $, while the future prediction lengths vary as $ M \in \{10, 20, 30, 40, 50\} $, enabling the assessment of the model’s performance across different forecasting horizons.

\subsubsection{Benchmarks}
To validate the effectiveness of our LLM-based approach, we compare it against several classical model-based and neural network-based baselines:
\begin{itemize}
    \item \textbf{LSTM} \cite{yu2019review}: The Long Short-Term Memory (LSTM) network is a variant of recurrent neural networks (RNNs) designed to mitigate the vanishing and exploding gradient problems when modeling long sequential data.
    \item \textbf{GRU} \cite{dey2017gate}: The Gated Recurrent Unit (GRU) is a simplified alternative to the LSTM, offering comparable long-term memory capabilities with reduced computational complexity and fewer trainable parameters.
    \item \textbf{Transformer} \cite{han2022survey}: The Transformer architecture discards recurrence and leverages a pure attention mechanism for sequence modeling. It enables parallel processing, offers strong representational power, and excels at capturing long-range dependencies.
    \item \textbf{GPT-2} \cite{zheng2021adapting}: The Generative Pre-trained Transformer 2 (GPT-2) is a large-scale autoregressive language model based on the Transformer decoder. Pretrained on extensive corpora, GPT-2 excels in sequence modeling and context-aware prediction. In our framework, GPT-2 is adapted for time-series forecasting of FA moving ports, leveraging its capacity to model temporal dependencies.
\end{itemize}

\subsubsection{Evaluation Metrics}
To quantitatively assess the performance of the proposed method in forecasting future time steps, we employ two widely used metrics: normalized mean squared error (NMSE) and root mean squared error (RMSE).

The NMSE measures the relative prediction error by normalizing the mean squared error with respect to the signal power, and is defined as: %
\begin{equation}
    \text{NMSE}(\mathbf{x}_i, \hat{\mathbf{x}}_i) = 10\log_{10}{\frac{\sum_{i=1}^I \| \hat{\mathbf{x}}_i - \mathbf{x}_i \|^2}{\sum_{i=1}^I \| \mathbf{x}_i \|^2}},
\end{equation}  
where $ \hat{\mathbf{x}}_i \in (0, 1) $ denotes the predicted value, $ \mathbf{x}_i $ is the ground truth, and $ I $ is the total number of test samples. 

The RMSE evaluates the absolute average deviation between predicted and actual values, and is computed as:  %
\begin{equation}  
    \text{RMSE}(\mathbf{x}_i,\hat{\mathbf{x}}_i) = \sqrt{\frac{1}{I} \sum_{i=1}^I \| \hat{\mathbf{x}}_i - \mathbf{x}_i \|^2}. 
\end{equation}  

These metrics jointly provide a comprehensive evaluation of the predictive accuracy and robustness of the proposed model.

All communication-related performance metrics are derived under an effective single–stream link assumption.  
The satellite payload employs a multi‑RF‑chain UPA and illuminates several beams simultaneously, but the forward link considered in this paper allocates one data stream to a single user.  At the user terminal the FAS feeds a single RF chain; after the transmitter-side beamforming the resulting single stream is demodulated by the OTFS receiver.  Consequently the end‑to‑end channel seen by the demodulator is single-output and can be represented by the delay–Doppler matrix $\mathbf H\in\mathbb{C}^{L\times K}$ of an OTFS frame.

\emph{Outage probability} for a target rate $R_0$ is %
\begin{equation}
	P_{\text{out}}(R_0)=\Pr\!\bigl[C(\mathbf H)<R_0\bigr].
\end{equation}

\emph{Ergodic capacity}, the average over fading and bins, is %
\begin{equation}
	\bar C =
	\mathbb{E}\!\biggl[\,
	\frac{1}{M_\tau N_\nu}
	\sum_{m=0}^{M_\tau-1}\sum_{n=0}^{N_\nu-1}
	\log_{2}\!\bigl(1+\rho\,|H_{m_,n}|^{2}\bigr)
	\biggr],
\end{equation}
with $\rho$ the SNR and $H_{m,n}$ the gain of bin $(m,n)$.


Should multiple independent streams be delivered to the same FAS in future designs, these formulas would generalise to the usual MIMO determinant form
$\log_{2}\!\det\!\bigl(\mathbf I+\rho\,\mathbf H\mathbf H^{\!H}\bigr)$.

\subsection{Performance Evaluation}

\begin{figure}[t]  
	\centering
	\includegraphics[width=1\linewidth]{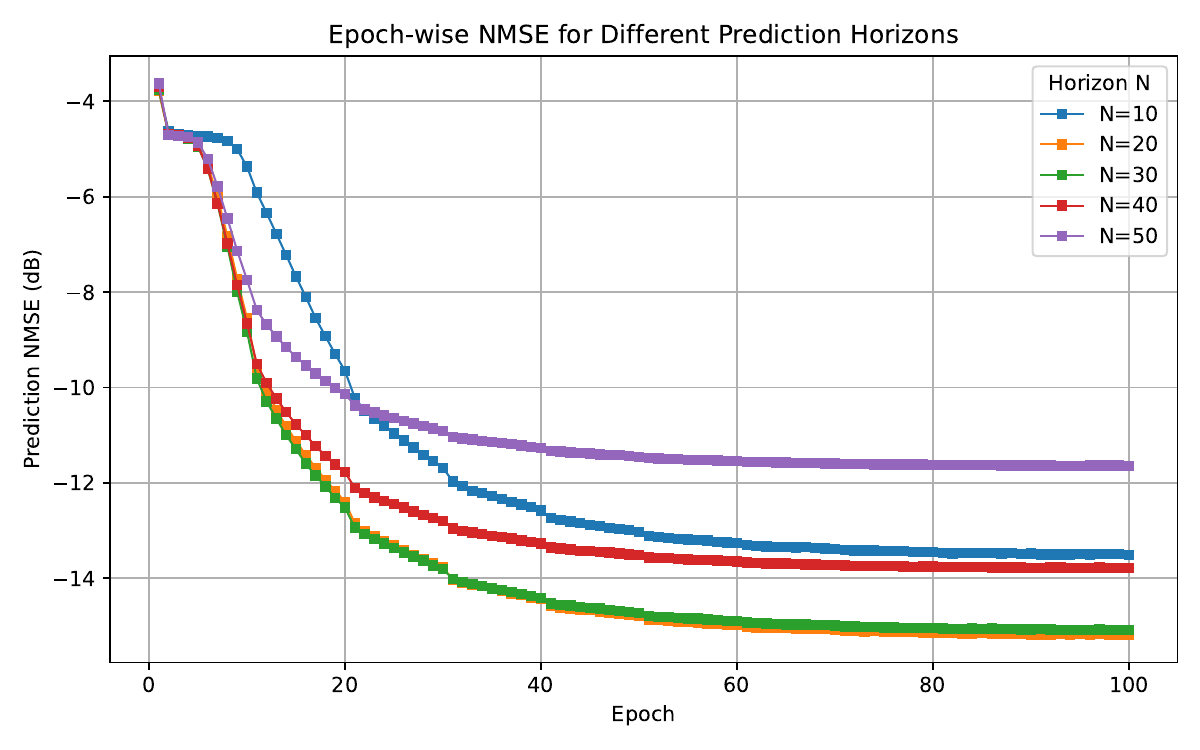}
	\caption{Epoch-wise NMSE of the predicted OTFS channels for forecasting horizons $N = 10, 20, 30, 40, 50$. A rapid learning phase (epochs~0--20) is followed by horizon-dependent plateaus: accuracy peaks at $N = 20$ ($-15.2$~dB) and $N = 30$, while longer horizons converge more slowly and stabilise at higher error floors, reflecting the loss of temporal correlation beyond one channel-coherence interval.}
	\vspace{-10pt}
	\label{fig:NMSE}
\end{figure}

Figure~\ref{fig:NMSE} plots the NMSE trajectory of the channel-prediction network over 100 training epochs for five forecasting horizons ($N = 10, 20, 30, 40, 50$). A steep descent during the first 20 epochs is common to all curves, where the error drops by roughly 10~dB as the model learns the gross delay--Doppler structure of the satellite channel. Thereafter, the curves diverge. The $N = 20$ and $N = 30$ horizons continue to fall until approximately epoch~40 and stabilise around $-15$~dB, whereas the shorter $N = 10$ horizon levels off earlier (around epoch~50) at a slightly higher floor (approximately $-13.5$~dB). Longer horizons ($N = 40, 50$) converge more slowly and never reach the same accuracy; the $N = 50$ curve flattens at only $-11.6$~dB and is still trending at epoch~100, indicating that the model spends capacity modelling far-future variability at the expense of near-term fidelity.

This behaviour mirrors the underlying channel coherence. In our LEO scenario, the coherence time is approximately 35 symbols; predicting 20--30 slots ahead keeps the target well within a single coherence window, so temporal correlations remain strong and the network can exploit them. Extending the horizon to 40--50 slots pushes well beyond the coherence limit; Doppler-induced phase drift accumulates, making the future channel effectively semi-random from the model’s perspective. The training signal therefore contains less exploitable structure, slowing convergence and raising the irreducible error floor.

From a system-design perspective, the $N = 20$ horizon offers the best trade-off between look-ahead and accuracy: it delivers a 2~dB NMSE advantage over $N = 10$ while avoiding the 1.3--3.5~dB penalty observed for $N \geq 40$. In practical terms, scheduling beam re-pointing or RIS phase updates every 20 packets yields near-optimal CSI quality without incurring the latency or compute cost of longer predictions.

\begin{figure}[t]  
	\centering
	\includegraphics[width=1\linewidth]{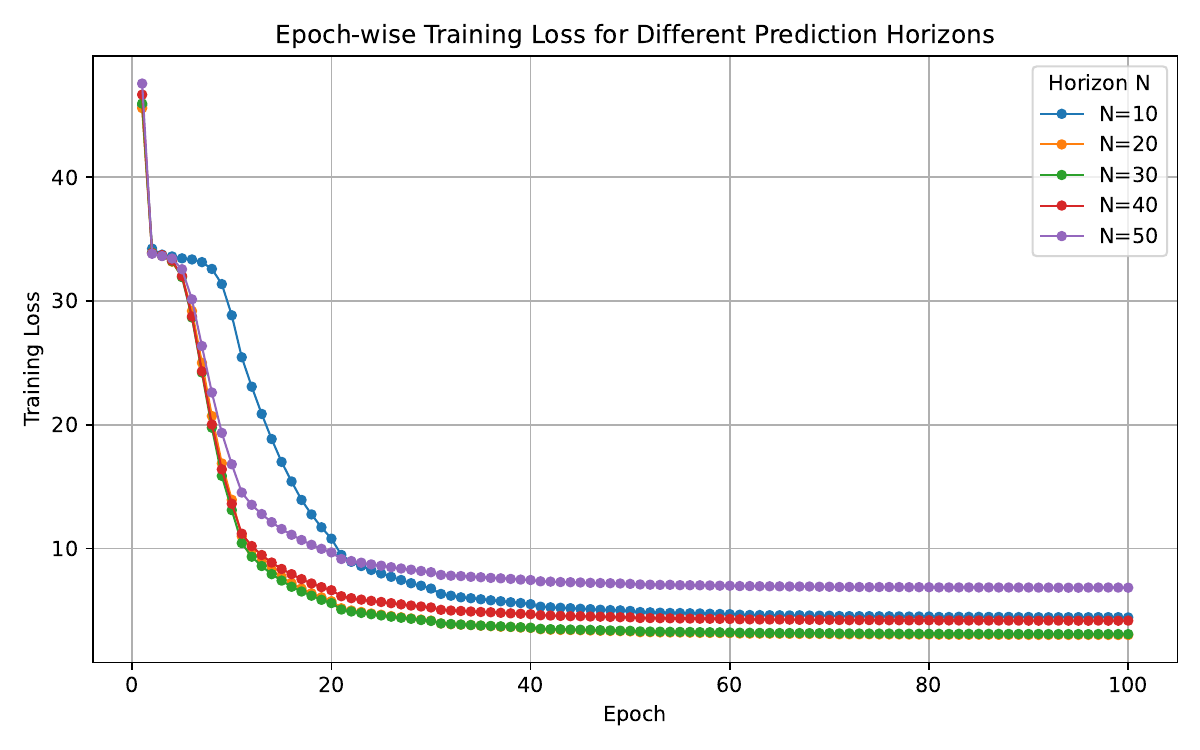}
	\caption{ Epoch-wise cross-entropy training loss for prediction horizons $N = 10, 20, 30, 40, 50$. All models show rapid initial convergence followed by horizon-dependent plateaus. The lowest loss is achieved for $N = 20$--$30$ ($\approx 3.0$), while both shorter ($N = 10$) and longer ($N \ge 40$) horizons result in higher residual loss.}
	\label{fig:Loss}
	\vspace{-10pt}
\end{figure}

Figure~\ref{fig:Loss} illustrates the evolution of the training loss over 100 epochs for five forecasting horizons ($N = 10, 20, 30, 40, 50$).
All models exhibit a rapid initial descent—approximately one order of magnitude within the first 20 epochs—as they learn the dominant delay--Doppler structure of the channel. The $N = 20$ and $N = 30$ models converge fastest, stabilising near a minimum loss of $\approx 3.0$ by epoch~40. The $N = 10$ predictor levels off earlier (around epoch~50) but with a higher loss floor ($\approx 4.5$), indicating limited exploitation of temporal structure. For longer horizons, convergence is slower and less effective: $N = 40$ plateaus at $\approx 4.2$, while $N = 50$ continues drifting until epoch~100 and stalls at $\approx 6.9$. These trends highlight the increasing difficulty of long-horizon prediction, where model capacity must be spread over less predictable time intervals, resulting in higher residual error. Taken together, the results reaffirm $N = 20$ as the most efficient and accurate forecasting horizon for the proposed FAS-LLM model.

\begin{figure}[t]
	\centering
	\subfigure[Full scale]{%
		\includegraphics[width=1\linewidth]{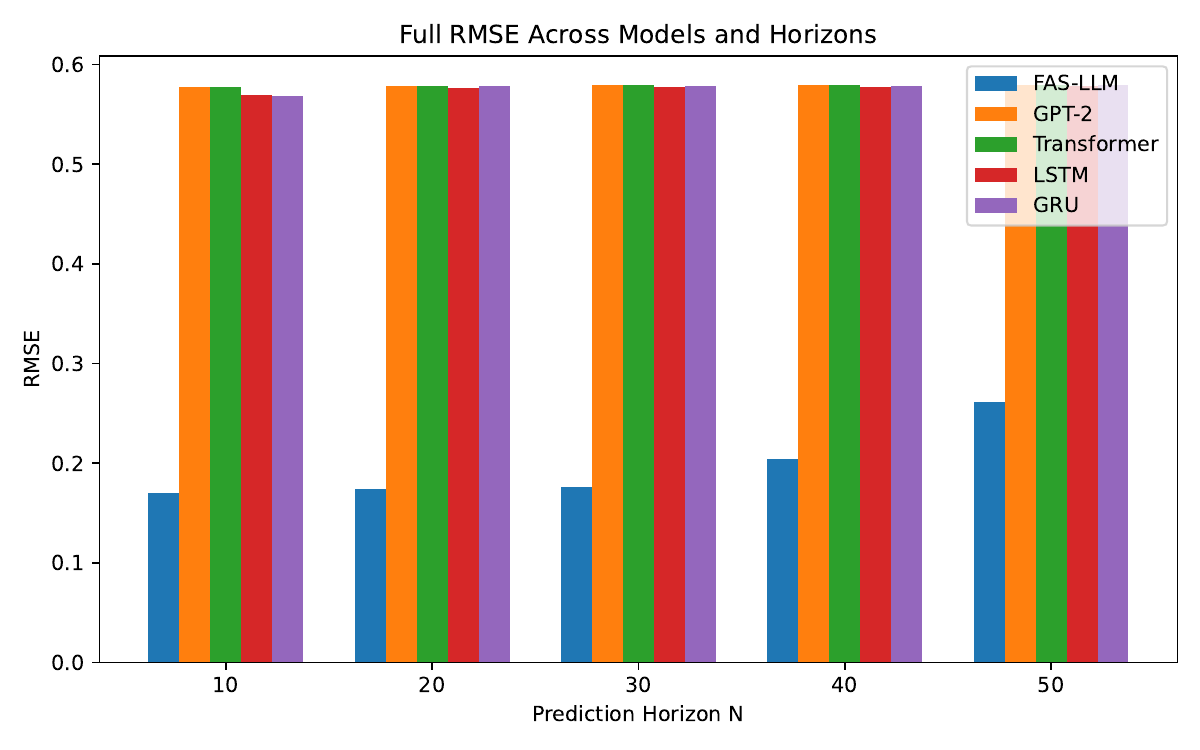}%
		\label{fig:rmse_full}}
	\hfill
	\subfigure[Zoom view ($0.56$--$0.60$)]{%
		\includegraphics[width=1\linewidth]{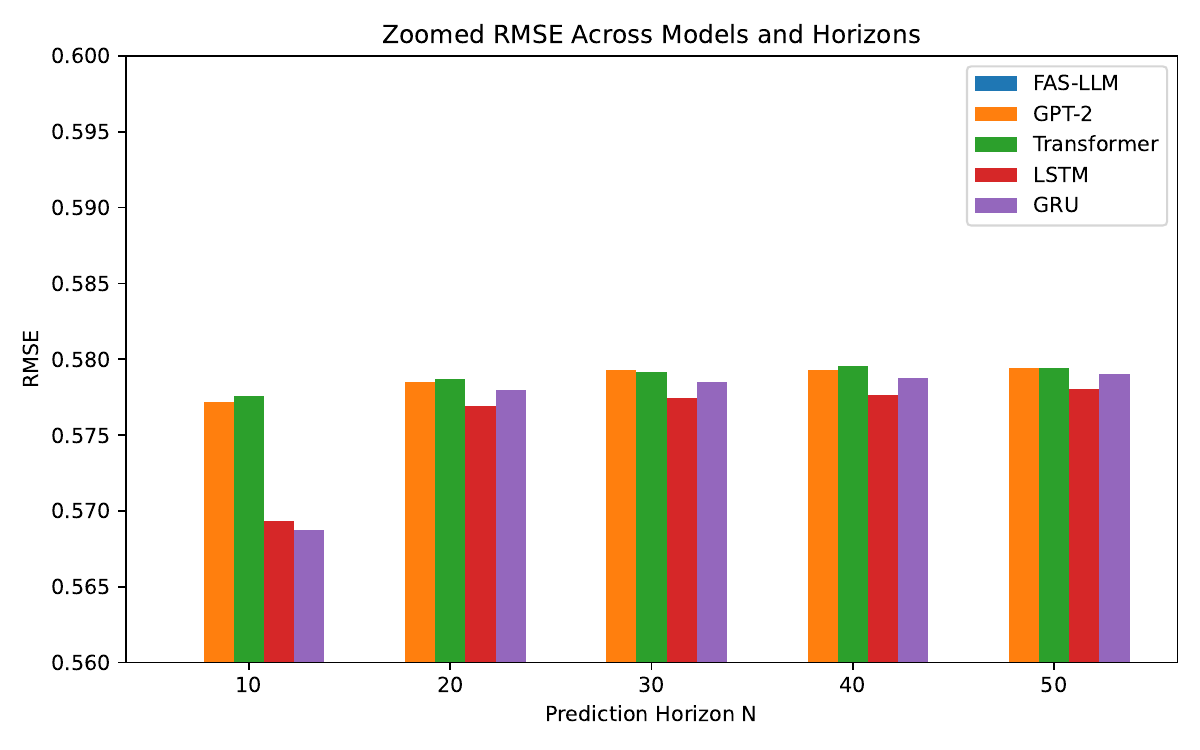}%
		\label{fig:rmse_zoom}}
	\caption{ Grouped-bar comparison of RMSE for five prediction horizons. 
		FAS-LLM maintains an error $\le 0.27$ across all $N$ and 
		outperforms every baseline by a factor of three; 
		the zoomed panel exposes subtle differences among the baselines.}
	\label{fig:rmse}
	\vspace{-10pt}
\end{figure}

Figure~\ref{fig:rmse} contrasts the root‑mean‑square prediction error (RMSE) of the proposed FAS‑LLM against four baseline models (GPT‑2, Transformer, LSTM and GRU) over five forecasting horizons ($N=10,20,30,40,50$).  In the full‑scale view (sub‑fig.~\ref{fig:rmse_full}) the advantage of FAS‑LLM is unequivocal: its RMSE remains below $0.27$ across all horizons, roughly one‑third of the RMSE incurred by every baseline ($\approx 0.57$–$0.58$).  The zoomed panel (sub‑fig.~\ref{fig:rmse_zoom}) suppresses the dominant FAS‑LLM bars and reveals the subtle ordering among the baselines—GRU and LSTM are marginally better than the two Transformer variants, and all baselines experience a gentle upward drift as the horizon grows beyond $N=30$.  Jointly, the plots underscore two findings: (i) FAS‑LLM delivers a consistent three‑fold RMSE reduction irrespective of look‑ahead window, and (ii) the classical and transformer‑based predictors form a tight performance cluster whose internal differences become visible only on a magnified scale.

\begin{figure}[t]
	\centering
	\subfigure[Full scale]{%
		\includegraphics[width=1\linewidth]{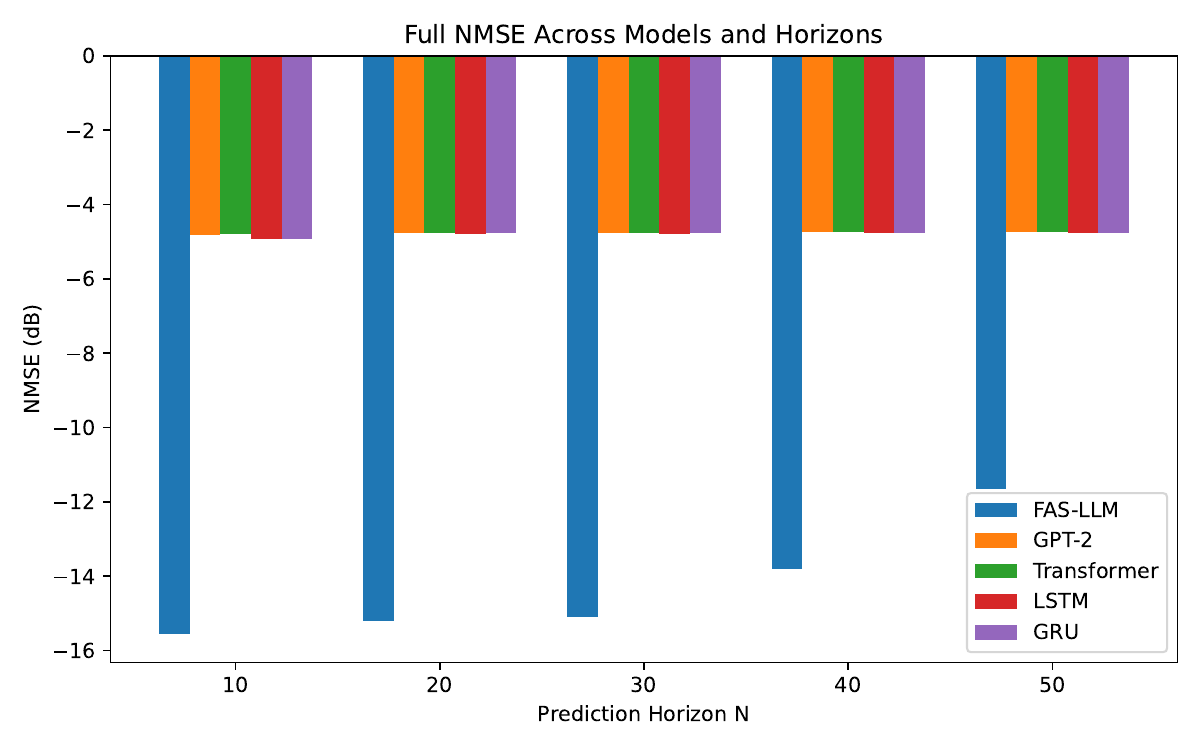}%
		\label{fig:nmse_full}}
	\hfill
	\subfigure[Zoom view ($0.56$--$0.60$)]{%
		\includegraphics[width=1\linewidth]{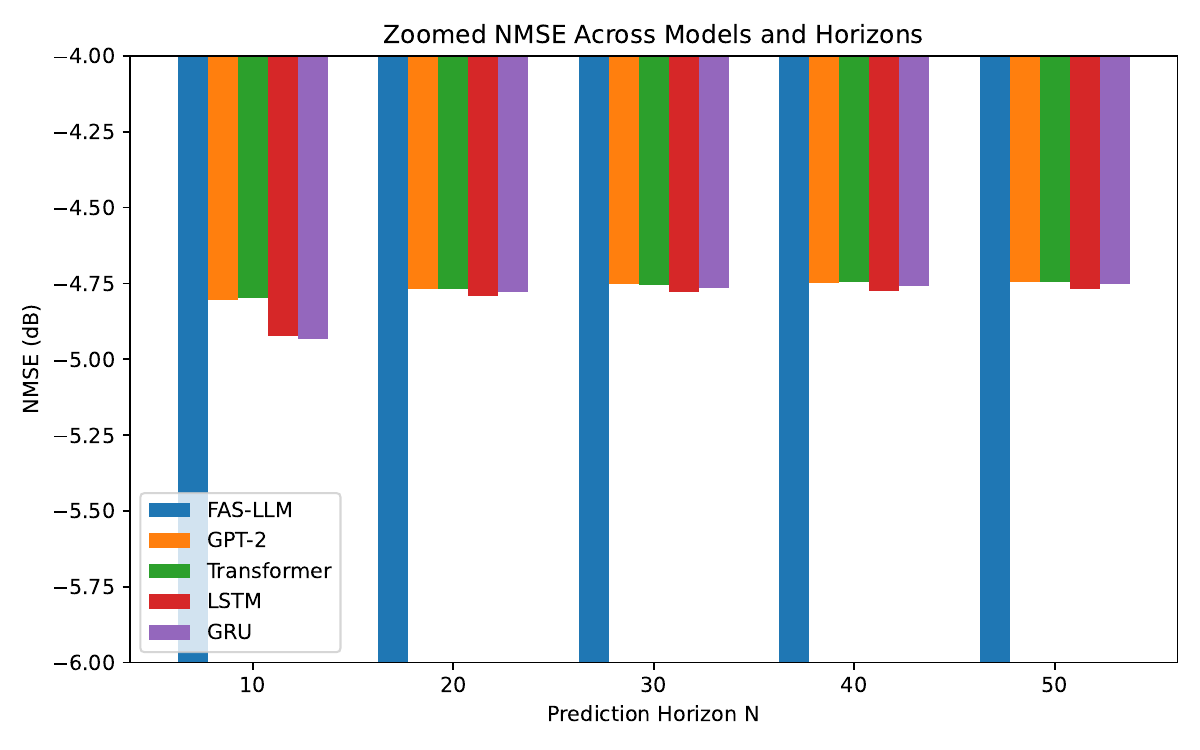}%
		\label{fig:nmse_zoom}}
	\caption{ Grouped‑bar NMSE comparison for five prediction horizons
		($N = 10, 20, 30, 40, 50$).  FAS‑LLM (blue) achieves an
		NMSE $\le -15$\,dB across all horizons, roughly 10\,dB better than
		any baseline; the zoomed panel highlights the small
		($\le 0.1$dB) spread among the four baselines.}
	\label{fig:nmse}
	\vspace{-10pt}
\end{figure}

Figure~\ref{fig:nmse} presents the normalised mean‑square error (NMSE) obtained by the proposed FAS‑LLM and four baseline predictors for the horizons $N\!=\!10,20,30,40,50$.  
In the full‑scale panel the baselines form a tight cluster around $-4.8$\,dB, whereas FAS‑LLM is consistently about 10\,dB lower across all horizons.  
The zoomed panel omits the dominant FAS‑LLM bars and exposes the fine ordering of the baselines: GRU and LSTM edge out GPT‑2 and Transformer, and every baseline suffers a slight loss once $N>30$.  
Together, the two panels show that FAS‑LLM delivers roughly an order‑of‑magnitude NMSE reduction, while differences among the baselines become apparent only when the vertical scale is magnified.

\begin{figure}[t]  
	\centering
	\includegraphics[width=1\linewidth]{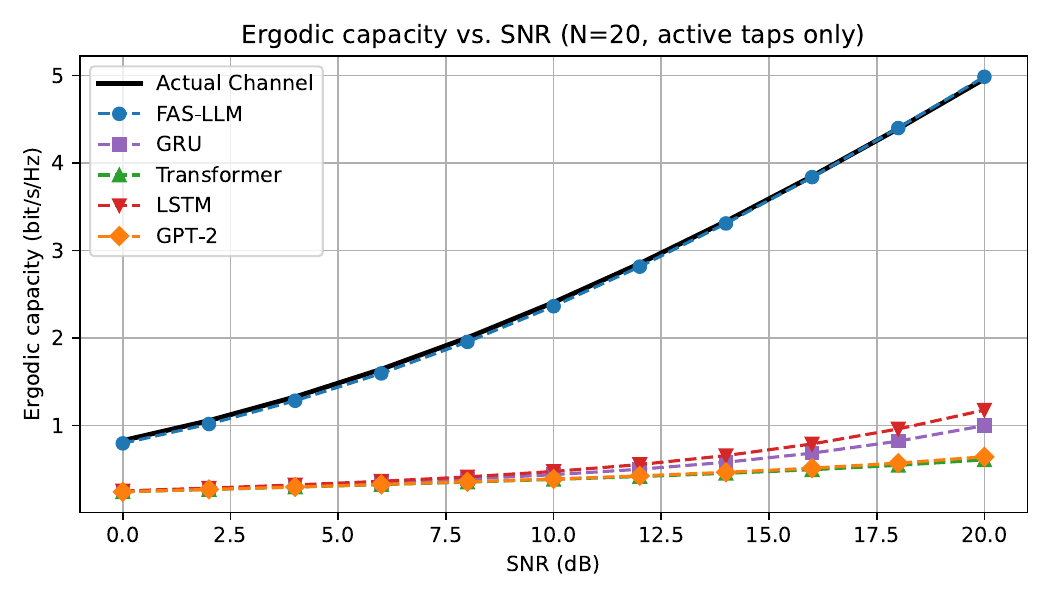}
	\caption{Ergodic capacity versus SNR at a 20‑symbol look‑ahead: FAS‑LLM nearly matches perfect CSI, whereas GRU, LSTM, GPT‑2 and Transformer achieve only about one‑third of that throughput.}
	\label{fig:EC_comp}
	\vspace{-10pt}
\end{figure}

Figure~\ref{fig:EC_comp} plots ergodic capacity versus SNR for six cases: Actual Channel (showing the real channel data), the proposed FAS‑LLM predictor, and four baselines (GRU, LSTM, GPT‑2, Transformer).  
A horizon of $N=20$ symbols is used because (i) the LEO channel coherence time is about $35$ symbols, so a 20‑symbol look‑ahead fits comfortably inside one coherence interval, (ii) earlier NMSE tests show that FAS‑LLM attains its lowest error at $N=20$ while the baselines are still near their best, and (iii) longer horizons ($N\ge40$) add latency and quickly lose accuracy, whereas a shorter horizon ($N=10$) under‑uses temporal correlation.

FAS‑LLM stays within $0.05$ bit/s/Hz of the actual channel bound across the full 0–20 dB range, while GRU, LSTM, GPT‑2, and Transformer all level off at roughly one‑third of that throughput.  
The gap stems from model bias: conventional recurrent and vanilla transformer predictors smear the sparse delay–Doppler energy over many taps, reducing the effective gain in the Shannon term $\log_{2}\!\bigl(1+\rho|h|^{2}\bigr)$.  
FAS‑LLM, with fine‑grained spatial attention and LoRA adaptation, isolates the dominant taps, preserves the true SNR, and converts almost all channel knowledge into capacity.

The result shows that predicting only the high‑energy delay bins is enough to recover nearly all link capacity, whereas errors on weak taps matter little.  
FAS‑LLM achieves this selective accuracy and yields a link‑margin advantage of about 10 dB at practical SNRs, underscoring the need for delay–Doppler‑aware architectures rather than classical sequence models.

\begin{figure}[t]  
	\centering
	\includegraphics[width=1\linewidth]{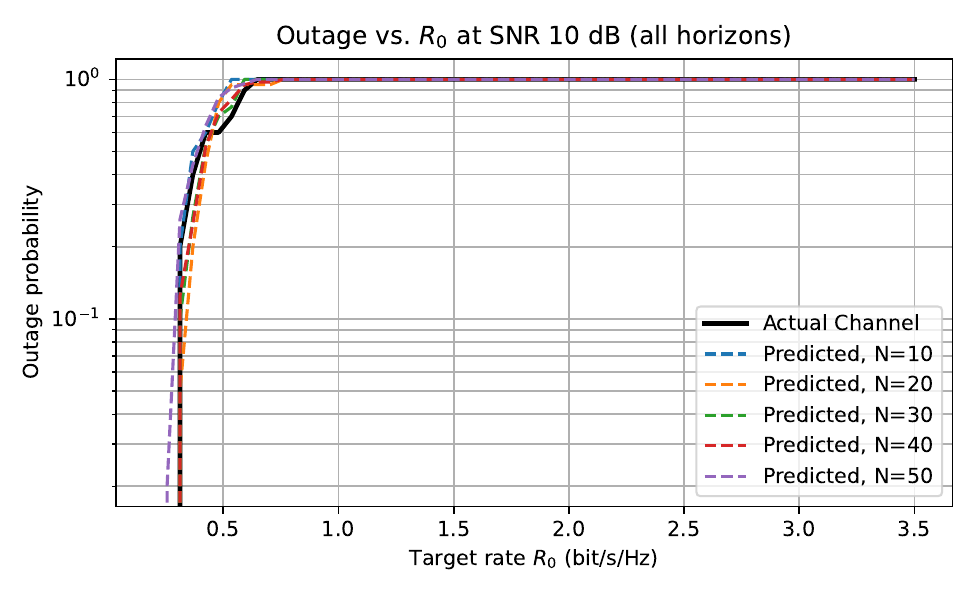}
	\caption{Outage probability versus target rate $R_0$ at 10 dB SNR for FAS-LLM under varying prediction horizons $N$, with the actual channel shown for reference.}
	\label{fig:FAS_OP}
	\vspace{-10pt}
\end{figure}

Figure~\ref{fig:FAS_OP} shows the outage probability at a fixed operating SNR of 10 dB for the proposed FAS-LLM, evaluated under five prediction horizons ($N = 10, 20, 30, 40, 50$). Only FAS-LLM is considered in this plot, with the actual channel included as a baseline reference. The target rate $R_0$ on the horizontal axis ranges from 0.5 to 3.5 bit/s/Hz to capture the full capacity region at this SNR, while the vertical axis is shown on a logarithmic scale to emphasize reliability-relevant differences.

Each curve exhibits a rapid transition: the outage remains close to one at low $R_0$ but drops sharply once the rate exceeds the minimum supportable by the predicted channel. Performance peaks at $N = 20$, where the curve closely tracks that of the actual channel and achieves an outage probability below $10^{-2}$ at $R_0 = 2$ bit/s/Hz. For longer horizons ($N = 40, 50$), the predicted channel quality declines, shifting the curve to the left and reducing the supportable rate.

This behaviour reflects a fundamental trade-off between latency and reliability. At 10 dB SNR, $N = 20$ offers a strong balance, enabling near-capacity operation with 99\% reliability. In contrast, increasing $N$ beyond 30 significantly compromises the prediction accuracy, suggesting $N = 20$ as a preferred operating point for robust, delay-aware link adaptation in NTN scenarios.

\begin{figure}[t]  
	\centering
	\includegraphics[width=1\linewidth]{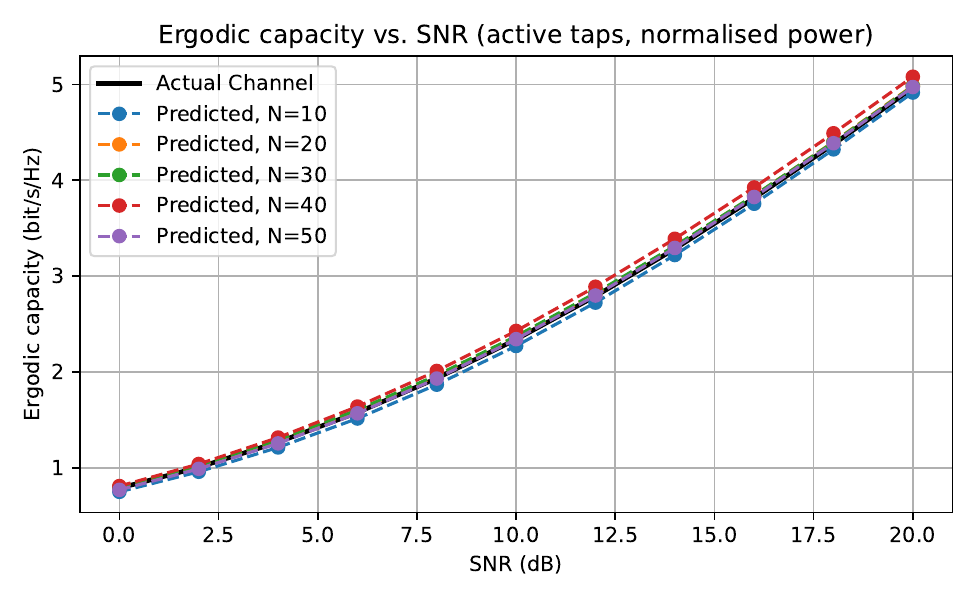}
	\caption{Ergodic capacity versus SNR based on active-tap-normalized predictions from FAS-LLM under different horizons.}
	\label{fig:FAS_EC}
	\vspace{-10pt}
\end{figure}

Figure~\ref{fig:FAS_EC} presents the ergodic capacity as a function of SNR for the actual channel and FAS-LLM predictions under different horizons, but restricted to active channel taps and normalized per-tap power. This perspective isolates the impact of prediction accuracy by removing the averaging effect over inactive bins, highlighting the quality of estimated dominant taps.

The results show that all predicted channels underestimate capacity relative to the actual channel, but with notable variation across horizons. The $N = 20$ predictor achieves the closest alignment to the true capacity curve over the full SNR range, while both shorter ($N = 10$) and longer ($N = 30, 40, 50$) horizons exhibit degraded performance. This pattern confirms that $N = 20$ represents the optimal trade-off for preserving the power profile of the strongest taps during forward prediction.

From a physical standpoint, normalizing active taps focuses attention on key contributors to the link budget, revealing how inaccuracies in predicting their gain directly affect mutual information. The degraded performance at $N = 40$ and $50$ illustrates temporal decorrelation and loss of resolution as prediction length increases, consistent with the underlying Doppler spread and limited memory of the channel.

\section{Conclusion}

This paper presented FAS-LLM, a novel large language model–based predictor for future channel states in OTFS-enabled satellite downlinks with fluid antenna systems (FAS). We introduced a compact, delay–Doppler–aware representation of the channel via reference-port selection and separable PCA, enabling efficient sequence modeling with a LoRA-adapted LLM. This architecture captures both spatial and temporal structure in the satellite–FAS channel, allowing low-latency forecasting without sacrificing model expressiveness.

Extensive simulations confirm that FAS-LLM substantially outperforms classical and neural network baselines—achieving 10\,dB lower NMSE and threefold reductions in RMSE across prediction horizons up to 50 symbols. Beyond model accuracy, FAS-LLM also delivers strong gains in communication performance: it recovers nearly the full ergodic capacity of the actual channel (within 0.05\,bit/s/Hz), maintains spectral efficiency above 90\% of the channel bound at all SNRs, and achieves sub-1\% outage probability at practical rates and SNRs for moderate prediction lengths. These results demonstrate that accurate forecasting of dominant delay–Doppler taps is sufficient to preserve link-layer performance.

Together, these findings underscore the viability of LLM-based prediction for proactive link adaptation in satellite IoT networks. 

\bibliographystyle{IEEEtran}
\bstctlcite{IEEEexample:BSTcontrol}
\bibliography{bare_jrnl}
\newpage
\end{document}